\documentstyle[12pt]{article}
\date {}
\input{pictex.sty}

\begin{document}

\centerline{{\bf BARYONS IN CHIRAL CONSTITUENT QUARK MODEL}}

\vspace{0.8cm}
\centerline{L. Ya. GLOZMAN}
\vspace{0.2cm}
\centerline{{\it Institut f\"ur Theoretische Physik,
Universit\"at Graz,}}
\centerline{{\it A-8010 Graz, Austria}}

\begin{abstract}
Beyond the spontaneous chiral
symmetry  breaking scale light and strange baryons
should be considered as  systems
of three constituent quarks with an effective confining interaction and a
flavor-spin chiral interaction that is mediated by the octet 
of Goldstone bosons
(pseudoscalar mesons) between the constituent quarks. One cannot 
exclude, however, the possibility that this flavor-spin interaction has an
appreciable vector- and higher meson exchange component.
\end{abstract}

\vspace{0.5cm}

\noindent
{\bf 1 Introduction}

\vspace{0.5cm}

Our aim in physics is not only to calculate some observable and get a
correct number but mainly to understand a physical picture responsible
for the given phenomenon. It  very often happens that a theory formulated
in terms of fundamental degrees of freedom cannot answer such a question
since it becomes overcomplicated at the related scale. Thus a main task in this
case is to select those degrees of freedom which are indeed essential.
For instance, the fundamental degrees of freedom in crystals are ions
in the lattice, electrons and the electromagnetic field. Nevertheless, in order
to understand electric conductivity, heat capacity, etc. we instead work
with  "heavy electrons" with dynamical mass, phonons and their interaction.
In this case a complicated electromagnetic interaction of the electrons with
the ions in the lattice is "hidden" in the dynamical mass of the electron
and the
interactions among ions in the lattice are eventually responsible  for the
 collective excitations of the lattice - phonons,
which are Goldstone bosons of the spontaneously broken translational
invariance in the lattice of ions.
As a result, the theory becomes rather
simple - only the electron and phonon degrees of freedom and their interactions
are essential for all the  properties of crystals mentioned above.

Quite a similar situation takes place in QCD. One hopes that sooner or later
one can solve the full nonquenched QCD on the lattice and get the
correct nucleon and
pion mass in terms of underlying degrees of freedom: current quarks and
gluon fields. However, QCD at the scale of 1 GeV becomes too complicated,
and hence it is rather difficult to say in this case what kind of
physics, inherent in
QCD, is relevant to the nucleon mass and its low-energy properties. 
I will show that it is the
spontaneous breaking of chiral
symmetry  which is the most important QCD phenomenon
in this case, and that beyond the scale of spontaneous breaking of chiral
symmetry light and strange baryons can be viewed as systems of three
constituent quarks which interact by the exchange of Goldstone bosons
(pseudoscaler mesons) and are subject to confinement 
\cite{GLO1,GLO2,GLO4,GPP,GPPVW}.\\

\noindent
{\bf 2 Spontaneous Chiral Symmetry  Breaking and its Consequences
for  Low-Energy QCD}

\vspace{0.5cm}

The QCD Lagrangian with three light flavors has a global symmetry

\begin{equation}SU(3)_{\rm L} \times SU(3)_{\rm R} \times U(1)_{\rm V} \times
U(1)_{\rm A},  \label{3.1} \end{equation}

\noindent
if one neglects the masses of current u,d, and s quarks, which are small
 compared to a typical  low-energy QCD scale of 1 GeV. The $U(1)_{\rm A}$
is not a symmetry at the quantum level due to the axial anomaly. If the
$SU(3)_{\rm L} \times SU(3)_{\rm R}$ chiral symmetry of the QCD Lagrangian were
intact
in the vacuum state we would observe  degenerate
multiplets in the particle spectrum corresponding
to the above chiral group,  and all hadrons would
have their degenerate partners with opposite parity. Since this does not
happen the implication is that the chiral symmetry is spontaneously
broken down to $SU(3)_{\rm V}$ in the QCD vacuum, i.e., realized in the hidden
Nambu-Goldstone mode. A direct evidence for the spontaneously broken
chiral symmetry is a nonzero value of the quark condensates for the
light flavors
$<{\rm vacuum}|\bar{q}q|{\rm vacuum}> \approx -(240-250 {\rm
MeV})^3$, 
which represents the order parameter. That this is indeed so, we know from
three independent sources: current algebra \cite{GOR},
QCD sum rules \cite{SHIF}, and lattice gauge calculations \cite{DANIEL}.
There are two important generic consequences of the spontaneous chiral symmetry
 breaking. The first one is an appearance of the octet
of pseudoscalar mesons of low mass, $\pi, {\rm K}, \eta$, which represent
the associated approximate Goldstone bosons. The second one is that valence
quarks acquire a dynamical or constituent mass.
Both these
 consequences of the spontaneous chiral symmetry  breaking
are well illustrated by, e.g. the $\sigma$-model \cite{LEVY} or the Nambu and
Jona-Lasinio model \cite{Nambu}.
We cannot say at the
moment for sure what the microscopical reason for spontaneous chiral symmetry
 breaking in the QCD vacuum is. It was suggested that this occurs
when quarks propagate through instantons in the QCD vacuum \cite{SHU,DIP}.

For the low-energy baryon properties it is only essential that
beyond the spontaneous chiral symmetry breaking scale  new
dynamical degrees of freedom appear - constituent quarks and chiral
fields \cite{MAG}. The low-energy baryon properties are mainly determined
by these dynamical degrees of freedom and the confining interaction.\\
 
\noindent
{\bf 3 The Chiral Boson Exchange Interaction}

\vspace{0.5cm}

In an effective chiral  description of the baryon
structure, based on the constituent quark model, the
coupling of the quarks and the pseudoscalar Goldstone
bosons will (in the $SU(3)_{\rm F}$ symmetric approximation) have
the form $ig\bar\psi\gamma_5\vec\lambda^{\rm F}
\cdot \vec\phi\psi$ (or
$g/(2m)\bar\psi\gamma_\mu
\gamma_5\vec\lambda^{\rm F}
\cdot \psi \partial^\mu\vec\phi$),
 where $\psi$ is the fermion constituent quark
field operator, $\vec\phi$ the octet boson field
operator, and $g$ is a coupling constant. A coupling of this
form, in a nonrelativistic reduction for the constituent quark spinors,
will -- to lowest order -- give rise to a Yukawa interaction
between the constituent quarks, the spin-spin component of which has
the form
\begin{equation}V_{\rm Y} (r_{ij})=
\frac{g^2}{4\pi}\frac{1}{3}\frac{1}{4m_im_j}
\vec\sigma_i\cdot\vec\sigma_j\vec\lambda_i^{\rm F}\cdot\vec\lambda_j^{\rm F}
\{\mu^2\frac{e^{-\mu r_{ij}}}{ r_{ij}}-4\pi\delta (\vec r_{ij})\}
.\label{5.1} \end{equation}
Here $m_i$ and $m_j$ denote the masses of the interacting quarks,
and $\mu$ that of the meson. There will also be an associated
tensor component, which is discussed in ref. \cite{GLO2}.

At short range the simple form (\ref{5.1}) of the chiral boson exchange
interaction cannot be expected to be realistic and should only
be taken to be suggestive.
Because of the finite spatial extent of both the constituent
quarks and the pseudoscalar mesons
 the delta function in (\ref{5.1}) should be replaced by a finite
function, with a range of 0.6-0.7 fm, as suggested
by the
spatial extent of the mesons.
In addition, the radial behaviour of the Yukawa
potential (\ref{5.1}) is valid only if the boson field
satisfies a linear Klein-Gordon equation. The implications of the underlying
chiral symmetry of QCD
for the effective chiral Lagrangian (which in fact is not known),
which contains
constituent quarks as well as boson fields,
are that these boson fields cannot be described by linear
equations near their source. For a clarification on this important issue
see \cite{SCHLADMING}.

{\it At this stage the proper procedure should be to avoid further specific
assumptions about the short range behavior of
$V(r)$ in
(\ref{5.5}),  to extract instead  the required matrix elements of it
from the baryon spectrum, and to reconstruct by this an approximate
radial form of $V(r)$.
The overall minus sign in the
effective chiral boson interaction in (\ref{5.5}) corresponds to that of the
short range term in the Yukawa interaction. It is the latter that is of crucial
importance in baryon physics.}

The flavor structure of the pseudoscalar octet exchange interaction
 between two quarks $i$ and $j$ should be understood as
follows
:
\begin{equation}
-V(r_{ij})  \vec {\lambda^{\rm F}_i} \cdot \vec {\lambda^{\rm F}_j}
\vec\sigma_i\cdot\vec\sigma_j \nonumber 
= 
-\left(\sum_{a=1}^3 V_{\pi}(r_{ij}) \lambda_i^a \lambda_j^a
+\sum_{a=4}^7 V_{\rm K}(r_{ij}) \lambda_i^a \lambda_j^a
+V_{\eta}(r_{ij}) \lambda_i^8 \lambda_j^8\right)
\vec\sigma_i\cdot\vec\sigma_j. \label{5.5} \end{equation}
The first term in (\ref{5.5}) represents the pion-exchange interaction,
which acts only between
light quarks. The second term represents the Kaon
exchange interaction,
which takes place in u-s and d-s pair states. The $\eta$-exchange,
which is represented by the third term, is allowed
in all quark pair states. 

The interaction (\ref{5.5}) should be contrasted with the gluon-exchange
one \cite{RGG}, which has been used in numerous earlier attempts to describe
the baryon spectra with the constituent quark model (see e.g. \cite{IGK1}). 
The gluon-exchange
model fails to explain the following outstanding features of baryon
spectroscopy: ({\it i}) the different ordering of lowest positive and negative
parity states in the spectra of nucleon and $\Lambda$-hyperon (the gluon
exchange is sensitive to the spin and colour degrees of freedom of quarks
only, so the N and $\Lambda$ spectra should be very similar as they differ 
only in their flavor content); ({\it ii}) the fact that some of the two-quantum
excitations of positive parity in all spectra (e.g. $N(1440)$, $\Delta(1600)$,
$\Lambda(1600)$, $\Sigma(1660)$) lie below the one-quantum excitations
of negative parity; ({\it iii}) an absense in the empirical spectra of the
strong spin-orbit splittings implied by the gluon-exchange interaction.
The gluon-exchange model is self-contradictory in principle: an introduction
into the theory of the constituent quark mass 
(to be contrasted with the small current quark mass)  implies that the
underlying chiral symmetry of QCD is spontaneously broken. If so, according
to the Goldstone theorem there must be Goldstone bosons which should
participate in the baryon structure on the same footing as constituent quarks.
\\

\noindent
{\bf 4 The Structure of the Baryon Spectrum}

\vspace{0.5cm}

The  two-quark matrix elements of the
interaction (\ref{5.5}) are:

$$<[f_{ij}]_{\rm F}\times [f_{ij}]_{\rm S} : [f_{ij}]_{\rm FS}
{}~| -V(r_{ij})\vec \lambda^{\rm F}_i \cdot \vec \lambda_j^{\rm F}
\vec\sigma_i \cdot \vec \sigma_j
{}~|~[f_{ij}]_{\rm F} \times [f_{ij}]_{\rm S} : [f_{ij}]_{\rm FS}> $$
\begin{equation}=\left\{\begin{array}{rl} -{4\over 3}V(r_{ij}) & [2]_{\rm
F},[2]_{\rm S}:[2]_{\rm FS} \\
-8V(r_{ij}) & [11]_{\rm F},[11]_{\rm S}:[2]_{\rm FS} \\
4V(r_{ij}) & [2]_{\rm F},[11]_{\rm S}:[11]_{\rm FS}\\ {8\over
3}V(r_{ij}) & [11]_{\rm F},[2]_{\rm S}:[11]_{\rm
FS}\end{array}\right..\label{6.1} \end{equation}

\noindent
{}From these the following important properties may be inferred:

(i) At short range, where $V(r_{ij})$ is positive, the chiral
interaction (\ref{5.5}) is attractive in the symmetric FS pairs and
repulsive in the antisymmetric ones. At large distances the potential
function $V(r_{ij})$ becomes negative and the situation is
reversed.

(ii) At short range,  among the FS-symmetrical pairs,
the flavor antisymmetric pairs experience
a much larger attractive interaction than the flavor-symmetric
ones, and among the FS-antisymmetric pairs
the strength of the repulsion in flavor-antisymmetric
pairs is considerably weaker than in the symmetric ones.

Given these properties we conclude, that
with the given flavor symmetry, the more symmetrical the FS
Young pattern is for a baryon the more attractive contribution at short
range comes from the interaction (\ref{5.5}). For two identical
flavor-spin Young patterns $[f]_{\rm FS}$ the attractive contribution
at short range is larger for  the more antisymmetrical
flavor Young pattern $[f]_{\rm F}$.

 Consider first, for the purposes of illustration, a schematic model
which neglects the radial dependence
of the potential function $V(r)$ in (\ref{5.5}), and assume a harmonic
confinement among quarks as well as $m_{\rm u}=m_{\rm d}=m_{\rm s}$.
In this model

\begin{equation}H_\chi\sim -\sum_{i<j}C_\chi~
\vec \lambda^{\rm F}_i \cdot \vec \lambda^{\rm F}_j\,
\vec
\sigma_i \cdot \vec \sigma_j.\label{6.2} \end{equation}

If the only interaction between the
quarks were the flavor- and spin - independent harmonic confining
interaction, the baryon spectrum would be organized in multiplets
of the symmetry group $SU(6)_{\rm FS} \times U(6)_{\rm conf}$. In this case
the baryon masses would be determined solely by the orbital structure,
and the spectrum would be organized in an {\it alternative sequence
of positive and negative parity states.}
The Hamiltonian (\ref{6.2}), within a first order perturbation theory,
 reduces the $SU(6)_{\rm FS} \times U(6)_{\rm conf}$ symmetry down to
 $SU(3)_{\rm F}\times SU(2)_{\rm S}\times U(6)_{\rm conf}$, which automatically
implies a splitting between the octet and decuplet baryons.

For the octet states ${\rm N}$, $\Lambda$, $\Sigma$,
$\Xi$ ($N=0$ shell, $N$ is the number of harmonic oscillator excitations
in a 3-quark state) as well as for their first
radial excitations of positive parity
 ${\rm N}(1440)$, $\Lambda(1600)$, $\Sigma(1660)$,
$\Xi(?)$ ($N=2$ shell) the flavor and spin symmetries
are $[3]_{\rm FS}[21]_{\rm F}[21]_{\rm S}$, and the contribution of the
Hamiltonian (\ref{6.2})
 is $-14C_\chi$. For the decuplet states
$\Delta$, $\Sigma(1385)$, $\Xi(1530)$, $\Omega$ ($N=0$ shell)
the flavor and spin symmetries,
as well as the corresponding matrix element, are
$[3]_{\rm FS}[3]_{\rm F}[3]_{\rm S}$
and $-4C_\chi$, respectively. The first negative parity excitations
($N=1$ shell) in the ${\rm N}$ and $\Sigma$ spectra ${\rm N}(1535)$ - ${\rm
N}(1520)$
and $\Sigma(1750)$ - $\Sigma(?)$
are described by the $[21]_{\rm FS}[21]_{\rm F}[21]_{\rm S}$ symmetries, and
the contribution
of the interaction (\ref{6.2}) in this case is $-2C_\chi$. The first negative
parity excitation in the $\Lambda$ spectrum ($N=1$ shell)
$\Lambda(1405)$ - $\Lambda(1520)$ is flavor singlet $[21]_{\rm FS}[111]_{\rm
F}[21]_{\rm S}$,
and, in this case, the corresponding matrix element is $-8C_\chi$.

These  matrix elements alone suffice to prove that
the ordering of the lowest positive and negative parity states
in the baryon spectrum will be correctly predicted by
the chiral boson exchange interaction (\ref{6.2}).
The constant $C_\chi$ may be determined from the
N$-\Delta$ splitting to be 29.3 MeV.
The oscillator
parameter $\hbar\omega$, which characterizes the
effective confining interaction,
may be determined as  one half of the mass differences between the
first excited
$\frac{1}{2}^+$ states and the ground states of the baryons,
which have the same flavor-spin, flavor and spin symmetries
(e.g. ${\rm N}(1440)$ - ${\rm N}$, $\Lambda(1600)$ - $\Lambda$, $\Sigma(1660)$
- $\Sigma$),
to be
$\hbar\omega \simeq 250$ MeV. Thus the two free parameters of this simple model
are fixed and we can  make now predictions.
In the ${\rm N}$ and $\Sigma$ sectors the mass
difference between the lowest
excited ${1\over 2}^+$ states (${\rm N}(1440)$ and $\Sigma(1660)$)
and ${1\over 2}^--{3\over 2}^-$ negative parity pairs
 (${\rm N}(1535)$ - ${\rm N}(1520)$ and $\Sigma(1750)$ - $\Sigma(?)$) will then
be
\begin{equation}{\rm N},\Sigma:\quad m({1\over 2}^+)-m({1\over 2}^--{3\over
2}^-)=250\, {\rm
MeV}-C_\chi(14-2)=-102\, {\rm MeV},\label{6.3} \end{equation}
whereas for the $\Lambda$ system ($\Lambda(1600)$,
$\Lambda(1405)$ - $\Lambda(1520)$) it should be

\begin{equation}\Lambda:\quad m({1\over 2}^+)-m({1\over 2}^--{3\over
2}^-)=250\, {\rm
MeV}-C_\chi(14-8)=74\, {\rm MeV}. \label{6.4} \end{equation}

This simple example shows how the chiral interaction (\ref{6.2})
provides different ordering of the lowest positive and negative parity excited
states in the spectra of the nucleon and
the $\Lambda$-hyperon. This is a direct
consequence of the symmetry properties of the boson-exchange interaction
discussed at the beginning of this section.
Namely, the $[3]_{\rm FS}$ state in the ${\rm N}(1440)$, $\Delta(1600)$
and
$\Sigma(1660)$ positive parity resonances from the $N=2$ band feels a
much stronger
attractive interaction than the mixed symmetry state $[21]_{\rm FS}$ in the
${\rm N}(1535)$ - ${\rm N}(1520)$,
$\Delta(1620)$ - $\Delta(1700)$
and $\Sigma(1750)$ -$\Sigma(?)$ resonances of negative parity ($N=1$ shell).
Consequently the masses of the
positive parity states ${\rm N}(1440)$, $\Delta(1600)$  and
$\Sigma(1660)$ are shifted
down relative to the other ones, which explains the reversal of
the otherwise expected "normal ordering".
The situation is different for $\Lambda(1405)$ - $\Lambda(1520)$
and
$\Lambda(1600)$, as the flavor state of  $\Lambda(1405)$ - $\Lambda(1520)$ is
totally antisymmetric. Because of this the
$\Lambda(1405)$ - $\Lambda(1520)$ gains an
attractive energy, which is
comparable to that of the $\Lambda(1600)$, and thus the ordering
suggested by the confining oscillator interaction is maintained.

Consider now, in addition, the radial dependence of the potential
with the $SU(3)_{\rm F}$ invariant version (\ref{5.5}) of
the chiral boson exchange interaction (i.e., $V_\pi (r)
=V_{\rm K} (r)=V_\eta(r)$).
If the confining interaction in each quark pair
is taken  to have the harmonic oscillator form as above,
the exact eigenvalues and eigenstates to
the coinfining 3q Hamiltonian
 are
$E=(N+3)\hbar\omega+3V_0$,
$\Psi=|N(\lambda\mu)L[f]_{\rm X}[f]_{\rm FS}[f]_{\rm F}[f]_{\rm
S}>$,
where $N$ is the number of quanta in the state, the Elliott symbol
$(\lambda \mu)$ characterizes the $SU(3)$ harmonic oscillator symmetry,
and $L$ is the orbital angular momentum. The spatial (X), flavor-spin (FS),
flavor (F), and spin (S) permutational symmetries are indicated
by corresponding Young patterns (diagrams) $[f]$. 


\begin{table}[!t]
\caption{The structure of the $\Lambda$-hyperon
states up to $N=2$, including predicted unobserved or nonconfirmed states,
indicated
by question marks. The predicted energies (in MeV)
are given in the brackets under the empirical values.}
{\footnotesize
\begin{center}
\begin{tabular}{|llll|} \hline
$ N (\lambda\mu)L[f]_{\rm X}[f]_{\rm FS}[f]_{\rm F}[f]_{\rm S}$
& LS multiplet & average &$\;\;\; \delta M_\chi$\\
&&energy&\\ \hline
$0(00)0[3]_{\rm X}[3]_{\rm FS}[21]_{\rm F}[21]_{\rm S}$ & ${1\over 2}^+,
\Lambda$ &
1115&$-14 P_{00}$\\
&&&\\
$1(10)1[21]_{\rm X}[21]_{\rm FS}[111]_{\rm F}[21]_{\rm S}$ & ${1\over 2}^-,
\Lambda(1405);
{3\over 2}^-,\Lambda(1520)$ &
1462&$-12 P_{00}+4P_{11}$\\
&&(1512)&\\
$2(20)0[3]_{\rm X}[3]_{\rm FS}[21]_{\rm F}[21]_{\rm S}$ & ${1\over 2}^+,
\Lambda(1600)$ &
1600&$-7 P_{00}-7P_{20}$\\
&&(1616)&\\
$1(10)1[21]_{\rm X}[21]_{\rm FS}[21]_{\rm F}[21]_{\rm S}$ & ${1\over 2}^-,
\Lambda(1670);
{3\over 2}^-, \Lambda(1690)$ &
1680&$-7 P_{00}+5 P_{11}$\\
&&(1703)&\\
$1(10)1[21]_{\rm X}[21]_{\rm FS}[21]_{\rm F}[3]_{\rm S}$ & ${1\over 2}^-,
\Lambda(1800);
{3\over 2}^-,\Lambda(?);$ &
1815&$-2 P_{00}+4P_{11}$\\
&${5\over 2}^-,\Lambda(1830)$&(1805)&\\
&&&\\
$2(20)0[21]_{\rm X}[21]_{\rm FS}[111]_{\rm F}[21]_{\rm S}$ & ${1\over 2}^+,
\Lambda(1810)
$&1810&$-6P_{00}-6P_{20}+4P_{11}$\\
&&(1829)&\\
$2(20)2[3]_{\rm X}[3]_{\rm FS}[21]_{\rm F}[21]_{\rm S}$ & ${3\over 2}^+,
\Lambda(1890);
{5\over 2}^+,\Lambda(1820)$ &
1855&$-7 P_{00}-7P_{22}$\\
&&(1878)&\\
$2(20)0[21]_{\rm X}[21]_{\rm FS}[21]_{\rm F}[21]_{\rm S}$&${1\over
2}^+,\Lambda(?)$&
?&$-{7\over 2}P_{00}-{7\over 2}P_{20}+5P_{11}$\\
&&(1954)&\\
$2(20)0[21]_{\rm X}[21]_{\rm FS}[21]_{\rm F}[3]_{\rm S}$ & ${3\over 2}^+,
\Lambda(?)$&
?&$-P_{00}-P_{20}+4P_{11}$\\
&&(1989)&\\
$2(20)2[21]_{\rm X}[21]_{\rm FS}[21]_{\rm F}[3]_{\rm S}$ & ${1\over 2}^+,
\Lambda(?);
{3\over 2}^+,\Lambda (?);$&2020?&
$-P_{00}-P_{22}+4P_{11}$\\
&${5\over 2}^+\Lambda(?);{7\over 2}^+,\Lambda(2020?)$&(2026)&\\
&&&\\
$2(20)2[21]_{\rm X}[21]_{\rm FS}[111]_{\rm F}[21]_{\rm S}$ & ${3\over 2}^+,
\Lambda(?);
{5\over 2}^+,\Lambda(?)$&
?&$-6P_{00}-6P_{22}+4P_{11}$\\
&&(2053)&\\
$2(20)2[21]_{\rm X}[21]_{\rm FS}[21]_{\rm F}[21]_{\rm S}$ & ${3\over
2}^+,\Lambda(?);
{5\over 2}^+,\Lambda(2110)$ &2110?
&$-{7\over 2}P_{00}-{7\over 2}P_{22}+5P_{11}$\\
&&(2085)&\\ \hline
\end{tabular}
\end{center}
}
\end{table}

When the
boson exchange interaction (\ref{5.5}) is treated in first order perturbation
theory, the mass of the baryon states takes the form
$M=M_0+N\hbar\omega+ \delta M_\chi$, 
where the chiral interaction contribution is
$ \delta M_\chi = <\Psi|H_\chi|\Psi>,$
and
$M_0 = \sum_{i=1}^3 {m_i} + 3(V_0 + \hbar \omega).$
The contribution from the chiral interaction  to each baryon
is a linear combination of the matrix elements of
the two-body potential $V(r_{12})$, defined as
$P_{nl}=<\varphi_{nlm}(\vec r_{12})
|V(r_{12})|\varphi_{nlm}(\vec r_{12})>$.

The  oscillator parameter $\hbar\omega$ and the four integrals
 are extracted from
the mass differences between the nucleon and the $\Delta(1232)$,
the $\Delta(1600)$ and
the ${\rm N}(1440)$, as well as the splittings between the nucleon
and the average mass of the
two pairs of states ${\rm N}(1535)-{\rm N}(1520)$ and ${\rm N}(1720)-{\rm
N}(1680)$.
This procedure yields the parameter values
$\hbar\omega$=157.4 MeV,
$P_{00}$=29.3 MeV, $P_{11}$=45.2 MeV, $P_{20}$=2.7 MeV and
$P_{22}$=--34.7 MeV. Given these values, all other excitation energies
(i.e., differences between the masses of given resonances and
the corresponding ground states)
of the nucleon, $\Delta$- and $\Lambda$-hyperon spectra are
predicted to within
 $\sim$ 15\% of the empirical values
where known, and are well within the uncertainty limits
of those values.
Note that these matrix elements provide a quantitatively satisfactory
description of the $\Lambda$-spectrum (Table 1)
even though they are extracted from the ${\rm N}-\Delta$ spectrum.\\


\noindent
{\bf 5 Three-Body Faddeev Calculations}

\vspace{0.5cm}

In the previous section we have shown how the Goldstone boson
exchange  (GBE), taken to first order
perturbation theory and without explicit parameterizing the radial dependence,
can explain the correct level ordering of positive and negative parity states
in light and strange baryon spectra, as well as the splittings in those
spectra.
A question, however, arises about what will happen beyond  first order
perturbation theory. In order to check this we have numericaly solved
three-body
Faddeev equations \cite{GPP, GPPVW}. Besides the confinement potential, which is
now taken in linear form, the GBE interaction between the constituent quarks
is now included to all orders. These results further support the adequacy
of the GBE for baryon spectroscopy.

In addition to the octet-exchange interaction we include here also the
flavor-singlet ($\eta'$) exchange. In the large $N_{\rm C}$ limit the axial
anomaly becomes suppressed \cite{WITT}, and the $\eta'$ becomes
the ninth Goldstone
boson of the spontaneously broken $U(3)_{\rm L} \times U(3)_{\rm R}$ chiral
symmetry.

We show our results  in  Fig. 1
It is well seen that the whole set of lowest ${\rm N}$ and $\Delta$
states is  reproduced quite correctly. In the most
unfavourable cases deviations  from the experimental
values do not exceed 3\%! In addition all level orderings are
correct. In particular, the positive-parity state ${\rm N}(1440)$ (Roper
resonance)
lies {\it below} the pair of negative-parity states ${\rm N}(1535)$ - ${\rm
N}(1520)$. The
same is true in the $\Delta$ spectrum with $\Delta(1600)$ and the pair
$\Delta(1620)$ - $\Delta(1700)$.

%
%
%
%

%
%
\begin{figure}[h]

%
%
\beginpicture

%
%
%
%
%
%
%
%
%

\setcoordinatesystem units <1.1mm,0.1mm>

%
%
\setplotarea x from 0 to 110, y from 900 to 1900

%
%
\axis left label {\stack{M,[MeV]}} ticks withvalues 
$900$ 
$1000$ 
$1100$ 
$1200$ 
$1300$ 
$1400$ 
$1500$ 
$1600$ 
$1700$ 
$1800$
 / 
at 900 1000 1100 1200 1300 1400 1500 1600 1700 1800 / /
%
{\Large
\axis bottom label {\Huge $\qquad N \qquad \qquad \qquad \quad \Delta$} 
ticks withvalues 
${\frac{1}{2}}^+$ 
${\frac{1}{2}}^-$
${\frac{3}{2}}^+$ 
${\frac{3}{2}}^-$ 
${\frac{5}{2}}^+$ 
${\frac{5}{2}}^-$ 
${\frac{1}{2}}^-$
${\frac{3}{2}}^+$ 
${\frac{3}{2}}^-$
/ 
at 10 20 30 40 50 60 80 90 100 / /
}
%
\shaderectangleson
\setshadegrid span <0.6mm>

%
%
%
\putrectangle corners at  7 1430 and  13 1470
\putrectangle corners at  7 1680 and  13 1740
\putrectangle corners at 17 1520 and  23 1555
\putrectangle corners at 17 1640 and  23 1680
\putrectangle corners at 27 1650 and  33 1750
\putrectangle corners at 37 1515 and  43 1530
\putrectangle corners at 37 1650 and  43 1750
\putrectangle corners at 47 1675 and  53 1690
\putrectangle corners at 57 1670 and  63 1685
\putrectangle corners at 77 1615 and  83 1675
\putrectangle corners at 87 1230 and  93 1235
\putrectangle corners at 87 1550 and  93 1700
\putrectangle corners at 97 1670 and 103 1770

\shaderectanglesoff

%
\linethickness0.3mm

%
\putrule from  7  940 to  13  940
\putrule from  7 1493 to  13 1493
\putrule from  7 1690 to  13 1690
\putrule from 17 1539 to  23 1539
\putrule from 17 1640 to  23 1640
\putrule from 27 1635 to  33 1635
\putrule from 37 1539 to  43 1539
\putrule from 37 1640 to  43 1640
\putrule from 47 1635 to  53 1635
\putrule from 57 1640 to  63 1640
\putrule from 77 1667 to  83 1667
\putrule from 87 1232 to  93 1232
\putrule from 87 1635 to  93 1635
\putrule from 97 1667 to 103 1667

%
\endpicture

\end{figure}

\vspace{1.5 cm}
\noindent
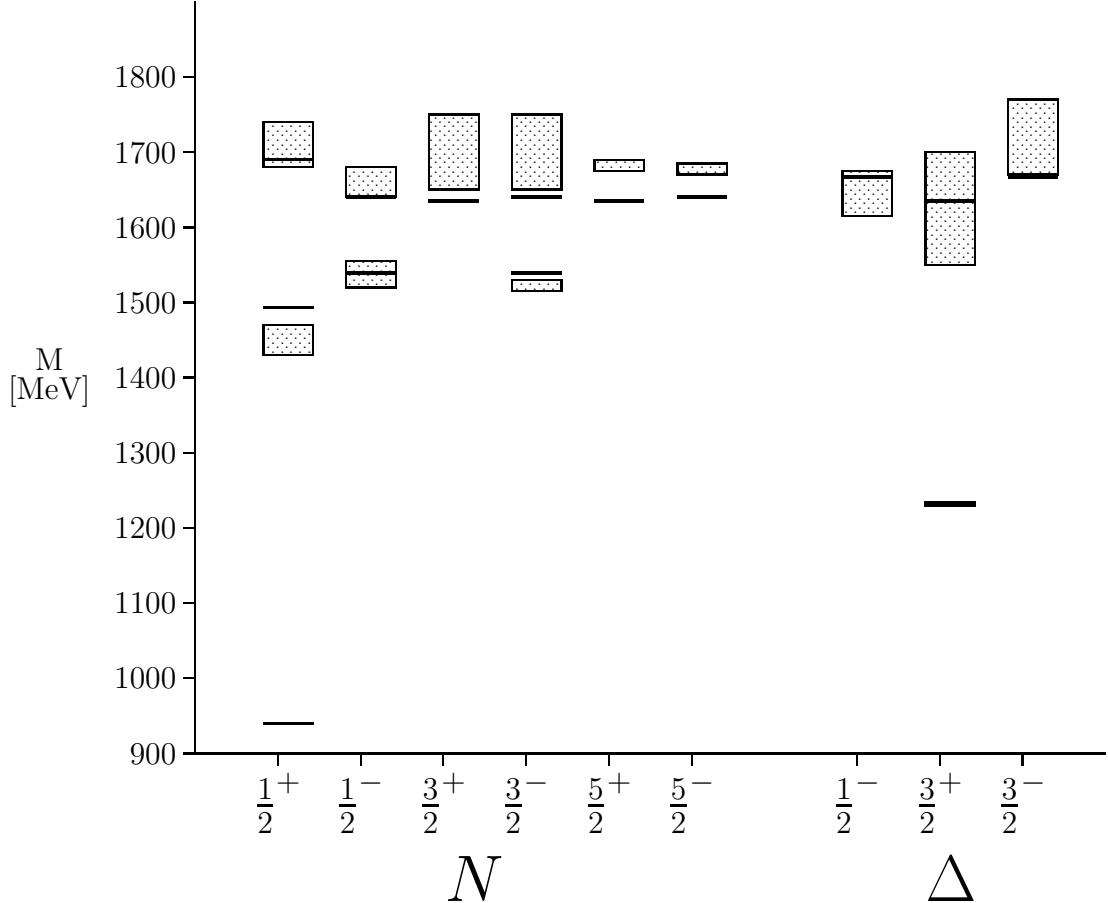
{\small Figure  1. Energy levels for the 14 lowest non-strange 
baryons with total
angular momentum and parity $J^P$. The shadowed boxes represent experimenta
uncertainties.}\\

\noindent
{\bf 6 Binding of Quarks and the $\pi N$ $\sigma$-Term}

\vspace{0.5cm}
The pion-nucleon $\sigma$-term is a measure of the explicit chiral symmetry 
breaking effects in the nucleon. The additive quark ansatz, where
the nucleon is considered as a system of three weakly interacting
constituent quarks,  leads to a much smaller value than the empirical result
extracted from pion-nucleon scattering data \cite{GASSER}. This indicates that
some essential piece of physics is absent within the additive quark ansatz.
It is shown in ref. \cite{SIGMA} that the contribution to $\sigma_{\pi N}$
that arises from the short range part of GBE between the constituent quarks is
crucial for the explanation of its empirical value.\\

\noindent
{\bf 7 Instead of a Conclusion}

\vspace{0.5cm}

Instead of a conclusion we discuss  some
important recent lattice QCD results in this last section.
It was shown already
a few years ago that one can obtain a qualitatively correct
splitting between $\Delta$ and ${\rm N}$ already within a quenched
approximation (for a review and references see \cite{WEINGARTEN}).
In the quenched approximation for baryons one takes into account
only 3 continuous valence quark lines and full gluodynamics.
This quenched approximation contains, however, part of antiquark
effects related to the Z graphs formed of valence quark lines.
One can even construct diagrams within the quenched approximation
which correspond to the exchange of the color-singlet isospin 1 or 0
${\rm q}\bar{\rm q}$ pairs between valence quark lines \cite{COHEN}. It is also
important that these diagrams contribute to the baryon mass to leading
order ($\sim N_{\rm C}$) in a $1/N_{\rm C}$ expansion  (their
contribution
to the $\Delta - {\rm N}$ splitting appears, however, to subleading orders).

{}From the quenched measurements \cite{WEINGARTEN} it is not clear
what were
the physical reason for the $\Delta - {\rm N}$ splitting:
gluon exchanges, instantons,
or something else. To clarify this question, Liu and Dong
have recently measured the $\Delta - {\rm N}$ splitting
in the quenched and a further so-called
"valence approximation" \cite{LIU}. In the valence approximation the
quarks are limited to propagating only forward in time (i.e., Z graphs
and related quark-antiquark pairs are removed). The gluon exchange and all
other possible gluon configurations, including instantons, are exactly the
same in both approximations. The striking result is that the $\Delta - {\rm N}$
splitting is observed only in the quenched approximation but not in the valence
approximation, in which the ${\rm N}$ and the $\Delta$ levels are degenerate
within
error bars. Consequently the $\Delta - {\rm N}$ splitting must receive a
considerable
contribution from  the diagrams with ${\rm q}\bar{\rm q}$ excitations, which
correspond
to the meson exchanges, but not from the gluon exchange or instanton-induced
interaction between quarks ( the instanton-induced interaction
could be rather important for the interactions between quarks and antiquarks
as it is  strongly attractive in the ${\rm q}\bar{\rm q}$ pseudoscalar
channel while it is  weak in ${\rm q}{\rm q}$ pairs).

Finally, the flavor-spin structure of the interaction (\ref{5.5}) is also
compatible with the vector meson ($\rho, K^*, \omega$) and axial-vector meson
exchanges \cite{DAN}. The radial behaviour of the spin-spin force associated
with the vector meson exchange
is similar to that of in (\ref{5.1}), while in the axial-vector
meson exchange case the contact term is absent and the required sign of the
interaction (\ref{5.5}) at short range comes from the Yukawa tail.
Thus one cannot exclude a possibility that in reality the  flavor-spin
interaction (\ref{5.5}) is some superposition of all possible meson
exchanges. It is well known that the spin-spin components of the 
pseudoscalar and vector meson exchange interactions have the same sign
while their tensor components tend to cancel. This could be an additional
reason for why the tensor force is not so important for baryons.\\

\noindent {\bf  Acknowledgements}\\

 It is a pleasure to thank Dan Riska, Zoltan Papp and Willi Plessas
for their collaboration that was crucial for the results presented here.


\begin{thebibliography}{99}
\bibitem{GLO1} L. Ya. Glozman and D. O. Riska, hep-ph/9411279; hep-ph/9412231.
\bibitem{GLO2}
L. Ya. Glozman and D. O. Riska,
Physics Reports {\bf 268} (1996) 263.
\bibitem{GLO4}  L. Ya. Glozman and D. O. Riska, Nucl. Phys. 
{\bf A 603} (1996) 326.
\bibitem{GPP}
L. Ya. Glozman, Z. Papp, and W. Plessas, Phys.  Lett. {\bf B381}
  (1996) 311.
\bibitem{GPPVW}
L. Ya. Glozman, Z. Papp, W. Plessas, K. Varga, R. Wagenbrunn, in preparation.
\bibitem{GOR}
M. Gell-Mann, R. J. Oakes, and B. Renner,
Phys. Rev. {\bf 175} (1968) 2195.
\bibitem{SHIF}
M. Shifman, A. Vainstein, and V. Zahkarov,
Nucl. Phys.  {\bf B147} (1979) 385, 448.
\bibitem{DANIEL}
D. R. Daniel et al, Phys. Rev. {\bf D46} (1992) 3130;
M. N. Fukugita et al, Phys. Rev. {\bf D47} (1993) 4739.
\bibitem{LEVY}
M. Gell-Mann and M. Levy,
 Nuovo Cim. {\bf 16} (1960) 705.
\bibitem{Nambu}
Y. Nambu and G. Jona-Lasinio, Phys. Rev. {\bf 122} (1961) 345.
\bibitem{SHU}
E.V. Shuryak,
Phys.Rep. {\bf C115} (1984) 152.
\bibitem{DIP}
 D. I. Diakonov and V. Yu. Petrov, Phys. Lett.
{\bf B147} (1984) 351; Nucl. Phys. {\bf B272} (1986) 457.
\bibitem{MAG}
A. Manohar and H. Georgi, Nucl. Phys. {\bf B234}
 (1984) 189.
\bibitem{SCHLADMING}
L.Ya. Glozman, Lecture given at the Schladming school on
Perturbative and Nonperturbative Aspects of
Quantum Field Theory (Schladming, Austria, March 1996), eds. H. Latal and
W. Schweiger, Springer, 1997 (hep-ph/9609278). 
\bibitem{RGG}
A. DeRujula, H. Georgi, and S. L. Glashow,
Phys. Rev. {\bf D12} (1975) 147.
\bibitem{IGK1}
N. Isgur and G. Karl, Phys. Rev. {\bf D18} (1978) 4187;
{\bf D19} (1979) 2653;
B. Silvestre-Brac and C. Gignoux,
Phys. Rev. {\bf D32} (1985) 743.
\bibitem{WITT}
E. Witten, Nucl. Phys.
 {\bf B156} (1979) 269.
\bibitem{GASSER}
J. Gasser, H. Leutwyler, and M. E. Sainio, Phys.  Lett. {\bf B253}
  (1991) 252.
\bibitem{SIGMA} L. Ya. Glozman, hep-ph/9608283.
\bibitem{WEINGARTEN}
D. Weingarten, Nucl. Phys. B (Proc. Suppl.) {\bf 34}
(1994) 29.
\bibitem{COHEN}
T. D. Cohen and D. B. Leinweber, Comments Nucl. Part.
Phys. {\bf 21} (1993) 137.
\bibitem{LIU}
K. F. Liu and S.-J. Dong,  hep-lat/9411067.
\bibitem{DAN}
K. Dannbom, L. Ya. Glozman, C. Helminen, D. O. Riska, hep-ph/9610384.


\end{thebibliography}
\end{document}